\documentclass[final,onecolumn]{elsarticle}
\usepackage{graphicx}
\usepackage{amssymb}
\usepackage{flushend,blindtext}
\usepackage[utf8]{inputenc}
\usepackage{enumitem}
\usepackage{amsmath, color}
\usepackage{amsmath}
\usepackage{amsthm}
\usepackage{subfig}
\usepackage{url}
\usepackage[backref=page, colorlinks=true,linkcolor=black, citecolor=<my_color>, urlcolor=blue]{hyperref}
\usepackage{lipsum}
\usepackage{mathtools}
\usepackage{cuted}
\usepackage[svganames]{xcolor}
\usepackage{float}
\usepackage{doi,xspace}
\usepackage[font=small]{caption}
\usepackage{graphicx}
\usepackage{lipsum}
\usepackage{multicol} 
\usepackage{booktabs}

\usepackage{hyperref}
\usepackage{cleveref}

\newcommand{\Graph}{\mathcal{G}}

\usepackage[linesnumbered,algoruled,boxed,lined]{algorithm2e}
\makeatletter 
\g@addto@macro{\@algocf@init}{\SetKwInOut{Parameter}{Parameters}} 
\makeatother

\newcommand{\subscr}[2]{#1_{\textup{#2}}}

\newcommand\ddfrac[2]{\frac{\displaystyle #1}{\displaystyle #2}}
\newcommand{\eps}{\epsilon}

\newcommand\oprocendsymbol{\hbox{$\square$}}
\newcommand\oprocend{\relax\ifmmode\else\unskip\hfill\fi\oprocendsymbol}

\usepackage[prependcaption,colorinlistoftodos]{todonotes}

\DeclareSymbolFont{bbold}{U}{bbold}{m}{n}
\DeclareSymbolFontAlphabet{\mathbbold}{bbold}
\newcommand{\vect}[1]{\mathbbold{#1}}
\newcommand{\vectorones}[1][]{\vect{1}_{#1}}
\newcommand{\vectorzeros}[1][]{\vect{0}_{#1}}

\def\prob{\mathbb{P}}

\def\diag{\operatorname{diag}}

\def\real{\mathbb{R}}

\newcommand{\lmax}{\lambda_{\max}}

\DeclareSymbolFont{bbold}{U}{bbold}{m}{n}
\DeclareSymbolFontAlphabet{\mathbbold}{bbold}
\biboptions{comma,round,authoryear}

\graphicspath{{./Figures/}}

\begin{document}
\begin{frontmatter}

\title{Multi-group connectivity structures and their implications\tnoteref{funding}} \tnotetext[funding]{This material is
  based upon work supported by, or in part by, the U.S.~Army Research
  Laboratory and the U.S.~Army Research Office under grant numbers
  W911NF-15-1-0577.}

\author[label1]{Shadi Mohagheghi}\corref{cor1}
\ead{shadi.mohagheghie@gmail.com}

\author[label2]{Pushkarini Agharkar}
\ead{pushkarini.a@gmail.com}

\author[label3]{Noah E. Friedkin}
\ead{friedkin@soc.ucsb.edu}

\author[label2]{Francesco Bullo} 
\ead{bullo@engineering.ucsb.edu}

\address[label1]{Department of Electrical and Computer Engineering, University of California at Santa Barbara, 
USA}

\address[label2]{Department of Mechanical Engineering, 
University of California at Santa Barbara, 
USA}

\address[label3]{Department of Sociology, University of California at Santa Barbara, USA.}

\cortext[cor1]{Corresponding author}


\begin{abstract} 
  We investigate the implications of different forms of
  multi-group connectivity. Four multi-group connectivity modalities are
  considered: co-memberships, edge bundles, bridges, and liaison
  hierarchies. We propose generative models to generate these four modalities. Our models are variants of planted partition or stochastic block models conditioned under certain topological constraints. We report findings
  of a comparative analysis in which we evaluate these structures, controlling for their edge densities and sizes, on mean
  rates of information propagation, convergence times to consensus, and
  steady state deviations from the consensus value in the presence of noise
  as network size increases. 
\end{abstract}

\end{frontmatter}

\section{Introduction}

\subsection{Motivation and problem description}
\begin{figure}[h]
  \begin{center} 
    \subfloat[Co-memberships]{\includegraphics[height=1.05in]{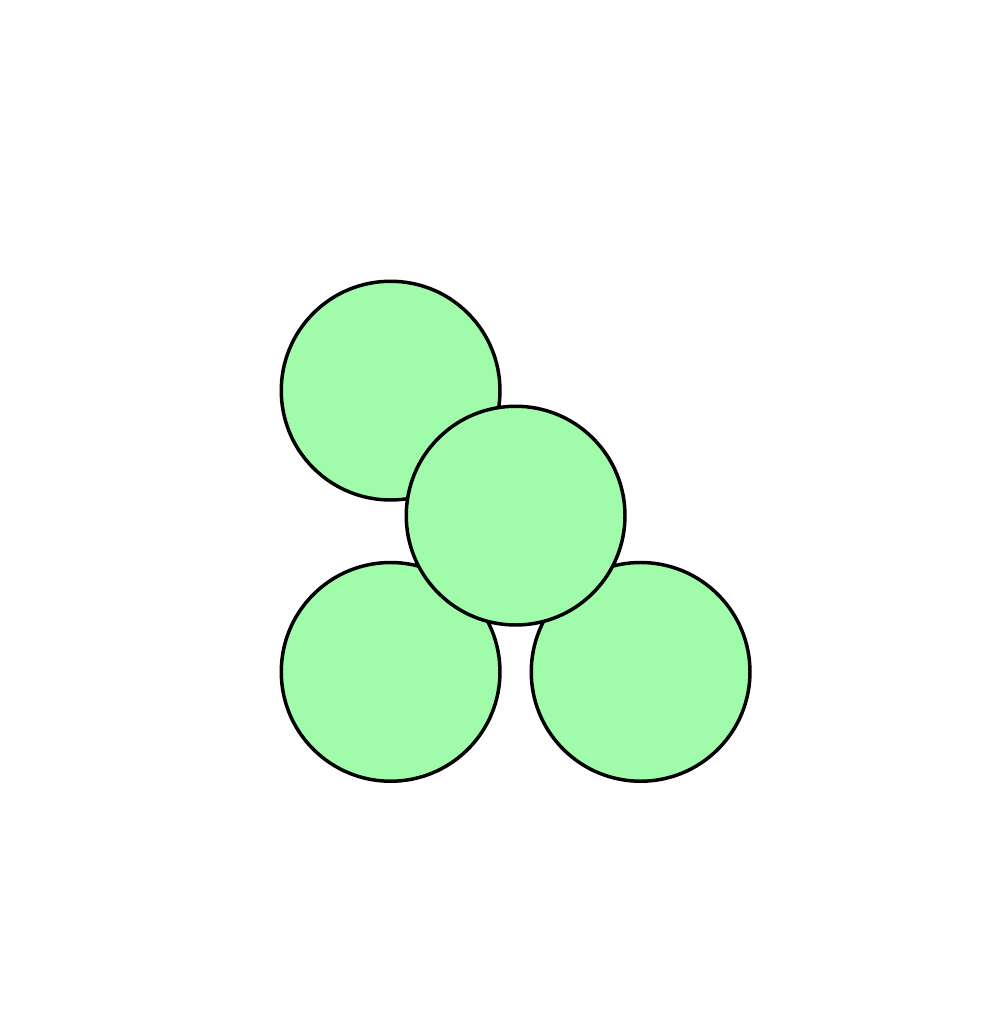}\label{fig:control-structure-3}}\qquad
    \subfloat[Edge Bundles]{\includegraphics[height=.95in]{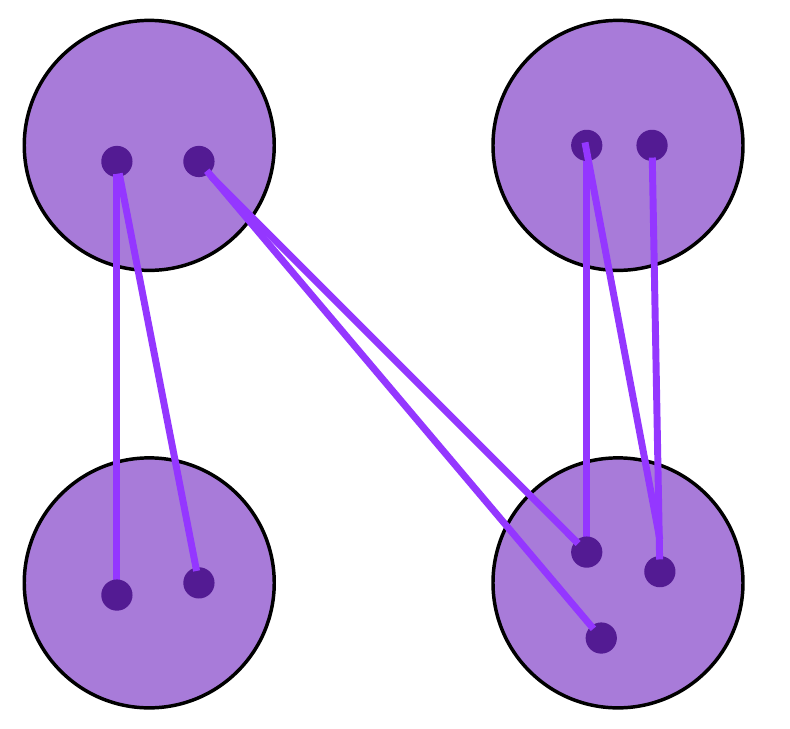}\label{fig:control-structure-2}}\\
    \subfloat[Bridges]{\includegraphics[height=.97in]{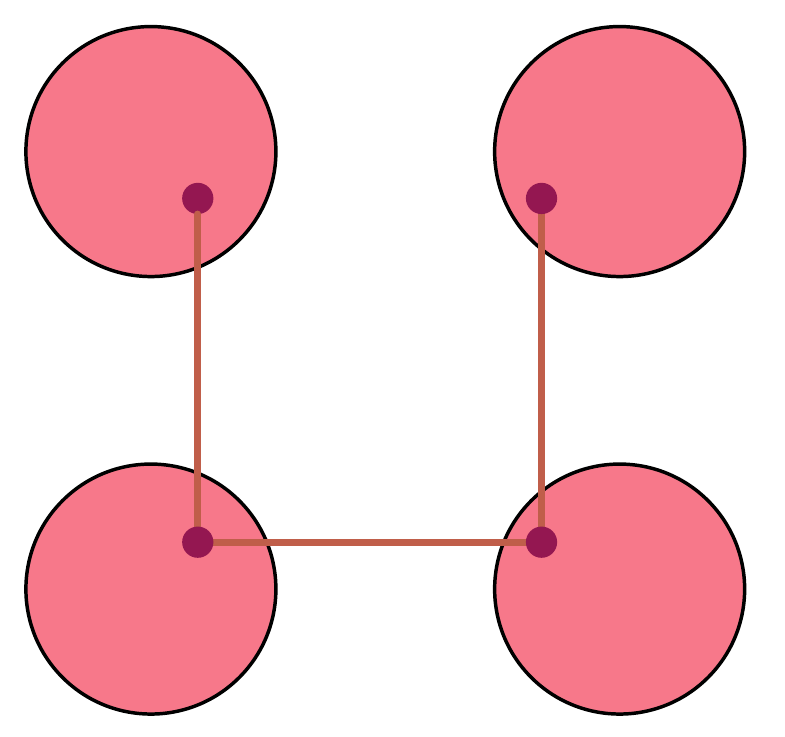}\label{fig:control-structure-1}}\qquad
    \subfloat[Liaisons]{\includegraphics[height=.975in]{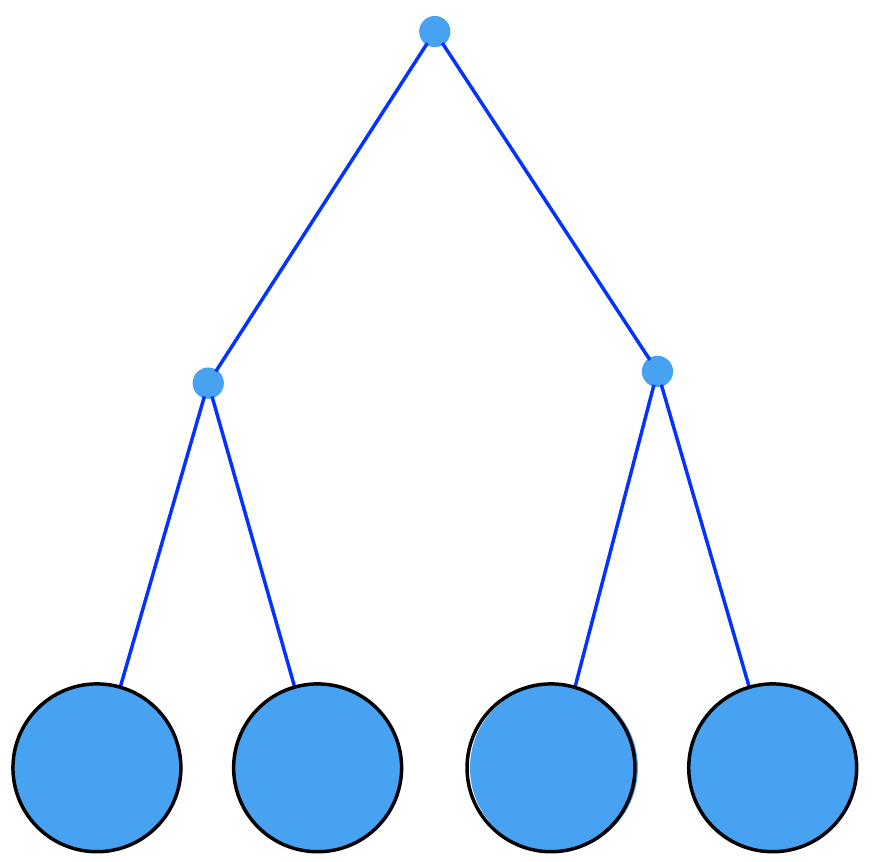}\label{fig:control-structure-4}}
    \caption{\small Small-scale illustration of the four forms of multi-group connectivity structures}\label{fig:control-structures}
  \end{center}
\end{figure}

As the size of a connected social network increases, multi-group formations
that are distinguishable clusters of individuals become a characteristic
and important feature of network topology. The connectivity of multi-group
networks may be based on co-memberships, edge bundles that connect multiple
individuals located in two disjoint groups, bridges that connect two
individuals in two disjoint groups, or liaison hierarchies of
nodes. Fig.~\ref{fig:control-structures} illustrates each form. A
large-scale network may include instances of all four connectivity
modalities. The work reported in this article is addressed to the
implications of these different forms of intergroup connectivity. We set up
populations of multiple subgroups and evaluate the implications of
different forms of intergroup connectivity structures. We analyze the
implications of different forms by adopting standard models of opinion
formation and information propagation that allow a comparative analysis on
metrics of mean rates of information propagation, convergence times to
consensus, and steady state deviations from the consensus value under
conditions of noise.

Typically, a corporation has formal hierarchical structure and additional
informal communication structures~\cite{RL:67}. The authority of the
large-scale organizations is subject to the well-known problem of control
loss, i.e., the cumulative decay of influence of superiors over
subordinates along the chain-of-command~\cite{OEW:70,NEF-ECJ:02}. Classic
and fascinating work on organization cultures~\cite{MC:64} points to the
importance of the topology of informal communication and influence networks
in mitigating and exacerbating coordination and control problems. Other
work has emphasized particular types of network typologies (linking-pin,
bridge, ridge, co-membership, and hierarchical) that may serve as structural
bases of mitigating coordination and control
loss~\cite{RL:67,NEF:98,MSG:73,DFS:77}. In this work we propose generative
network models and provide a comparative analysis for these typologies,
which we believe are lacking in the literature. Among the multitude of
possible coordination and control structures for large groups, we study
four prototypical structures and corresponding taxonomy shown in
Fig.~\ref{fig:control-structures}.  Bridge connected structure, in which
communication between subgroups are based on single contact edges between
subgroups; the coordination and control importance of such bridges is the
emphasis of the~\cite{MSG:73} model. According to Granovetter~\citep{MSG:73} only weak
ties can be bridges and those weak ties are more likely to be sources of
novel information making them surprisingly valuable. Additional references
include~\cite{MT-DK:10,WS-TE:08,MG:83,WRE-WDD:05}. Ridge connected or
redundant ties structure, in which multiple redundant contact edges connect
pairs of groups providing a robust basis of subgroup connectivity; the
coordination and control importance of such ridges is the emphasis
of~\cite{NEF:98}, Chapter~8.  Additional references
are~\cite{NEF:83,HCW-SAB-RLB:76,SAB-HCW:76}.  Co-membership intersection
structures, in which subgroups have common members; the coordination and
control importance of such structures is the emphasis of the linking-pin
model by Likert~\citep{RL:67}. This structure represents an organization as a
number of overlapping work units in which a member of a unit can belong to
other units. Further references
include~\cite{ENS-MSP-LANA:90,BC-JAH:04,SPB-DSH:11}.  Hierarchical
connected structure, in which distinct subgroups communicate through
liaisons, e.g., a star configuration in which a single individual (who may
or may not be in a command role) monitors and facilitates all the work by
subgroups and is responsible for all communications among them. Further
references are~\cite{JRG:74,EVR-JDJ:82,DFS:77,AVS-AJ-AG:14}.

We relate the generative models for the first three connectivity structures (co-memberships, edge bundles, and bridges) to stochastic block models (SBMs), which were first introduced in statistical sociology by Holland et al.~\citep{PWH-KL-SL:83} and Fienberg \& Wasserman~\citep{SEF-SSW:81}. Also known as planted partition model in theoretical computer science, SBM is a generative graph model that leads to networks with clusters. Conventionally, SBMs are defined for undirected binary graphs and non-overlapping communities. Generalizations of these models to digraphs~\cite{YJW-GYW:87}, overlapping memberships~\cite{EMA-DMB-SEF-EPX:08}, weighted graphs~\cite{CA-AZJ-AC:14} and arbitrary degree distributions~\cite{BK-MEJN:11} have also been studied.

In the field of social network science, the four forms of subgroup
connectivity illustrated in Fig.~\ref{fig:control-structures} are familiar
constructs. Comparative research on their implications is
limited. Granovetter~\citep{MSG:73} and Watts \& Strogatz~\citep{DJW-SHS:98} have focused on the
implications of multi-group connectivity based on bridges. Friedkin~\citep{NEF:98}
focused on co-membership and edge-bundle connectivity constructs, referring
to them as ``ridge'' structures. Reynolds \& Johnson~\citep{EVR-JDJ:82} focused on the
importance of liaisons. It may be that ridge structures provide a more
robust basis of influence and information flows than thinly dispersed
bridges and liaisons. We are unaware of any comparative analysis of all
four forms of inter-group connectivity structures that employs a common set
of dynamical-system behavioral metrics.

\subsection{Statement of contribution} 

In this article, we develop generative-network models that set up sample
networks for each form of multi-group connectivity topology and conduct a
comparative analysis of them, which we believe is lacking in the
literature. Our models, under some additional constraints, can be regarded as stochastic block models. We compare these network topologies on three metrics: (i)
spectral radius, that is a metric of the rate of information propagation in
a network propagation models, (ii) convergence time to consensus based on
the classic French-DeGroot opinion dynamics, and (iii) steady state
deviation from the French-DeGroot consensus value in the presence of
noise. We perform a regression analysis to obtain an equitable comparison on the performance of these four connectivity structures and to account for the discrepancies among their structural properties. We learned that the development of generative-network models,
suitable for this comparative analysis, is non-trivial. We lay out in
detail the assumptions of our models. This is the methodological
contribution of the article. The comparative analysis of network metrics,
over samples of networks of increasing size in the class of each form of
multi-group connectivity, is the article's theoretical contribution to a
better understanding of the implications of these different forms.

For network propagation processes, we refer to the classic
references~\cite{AL-JAY:76,HWH:78,LJSA:94} and to the recent
review~\cite{WM-SM-SZ-FB:16f}.  For opinion dynamic processes and the
French-DeGroot model, we refer to the classic
references~\cite{JRPF:56,MHDG:74} and the
books~\cite{NEF:98,MOJ:10,FB:18}.

\subsection{Preliminaries}

\paragraph{Graph theory}
Each graph $\mathcal{G(V, E)}$ is identified with the pair $\mathcal{(V,
  E)}$. The set of graph nodes $\mathcal{V} \neq \emptyset$ represents
actors or groups of actors in a social network. $|\mathcal{V}|=n$ is the
size of the network. The set of graph links $\mathcal{E}$ represents the
social interactions or ties among those actors. We denote the set of
neighbors of node $i$ with $\mathcal{N}_i$. In a weighted graph, edge
weights represent the frequency or the strength of contact between two
individuals, whereas in a binary graph all edge weights are equal to
one. The density of $\Graph$ is given by ratio of the number of its
observed to possible edges, $ \ddfrac{2|\mathcal{E}|}{n(n-1)}$. Graph
$\Graph$ is called dense if $|\mathcal{E}| = \mathcal{O}(n^2)$ and
sparse if $|\mathcal{E}| \ll n^2$.  A graph with density of 1 is a
\emph{clique}.

A walk of minimum length between two nodes is the shortest path or
geodesic. Average geodesic length is defined by $L=\ddfrac{1}{n(n-1)}
\sum_{i, j \in \mathcal{V}, i \neq j} d_{ij}$, where $d_{ij}$ is the length
of the geodesic from node $i$ to node $j$. A connected acyclic subgraph of
$\Graph$ spanning all of its nodes is a spanning tree. A uniform spanning
tree of size $n$ is a spanning tree chosen uniformly at random in the set
of all possible spanning trees of size $n$. Degree or connectivity of node
$i$ is defined as the number of edges incident on it.  The degree
distribution of a graph $P(k)$ is the number of nodes with degree $k$, or
the probability that a node chosen uniformly at random has degree $k$. The
clustering coefficient of node $i$ is given by the ratio of existing edges
between the neighbors of node $i$ over all the possible edges among those
neighbors. Letting $c_i= \ddfrac{2 e_{jk} : v_j, v_k \in \mathcal{N}_i,
  e_{jk} \in \mathcal{E}}{k_i(k_i -1 )}, k_i=|\mathcal{N}_i|$, the average
clustering coefficient of graph $\Graph$ is defined as $C= \frac{1}{n}
\sum_{i \in \mathcal{V}} c_i$.
  
An Erd\H{o}s-R\'{e}nyi graph~\cite{PE-AR:59} is constructed by connecting nodes randomly. Each edge is included in the graph with a fixed probability $p$ independent from every other edge. We represent such graph as
$\Graph_{ER}(n,p)$ where $p$ is the probability that each edge is
included in the graph independent from every other edge.  The probability
distribution of $\Graph_{ER}(n,p)$ follows a binomial distribution
$P(k) = {\binom{n-1}{k}} p^k (1-p)^{n-1-k}$, and its average clustering
coefficient is given as $C=p$.

\paragraph{Linear algebra} We denote the adjacency matrix of $\Graph$
with $A \in \mathbb{R}^{n \times n}$ whose $a_{ij} $th entry is equal to
the weight of the link between nodes $i$ and $j$ when such an edge exists,
and zero otherwise. Matrix $A$ is irreducible if the underlying digraph is
strongly connected. If digraph $\Graph$ is aperiodic and irreducible,
then $A$ is primitive. (A digraph is aperiodic if the greatest common
divisor of all cycle lengths is $1$.) A cycle is a closed walk, of at least
three nodes, in which no edge is repeated.

We adopt the shorthand notations $\vectorones[n] = [1,\dots,1]^{\top}$
and $\vectorzeros[n] = [0,\dots,0]^{\top}$. Given $x =
[x_1,\dots,x_n]^{\top} \in \real_n$, $\diag(x)$ denotes the diagonal
matrix whose diagonal entries are $x_1,\dots,x_n$. For an irreducible
nonnegative matrix $A$, $\lmax$ denotes the dominant eigenvalue of $A$
which is equal to the spectral radius of $A$, $\rho(A)$. The left positive eigenvector of $A$ associated with $\lmax$ is called the left dominant eigenvector of $A$.

\paragraph{Empirical networks properties} Our generative-network models attend to three often observed properties of real networks. (i) \textit{Small average shortest path}: in networks with a large number of vertices, the average shortest path lengths are relatively small due to the existence of bridges or shortcuts. (ii) \textit{Heavy tail degree distribution}: in contrast to Erd\H{o}s-R\'{e}nyi graphs with binomial degree
distribution, degree distributions of more realistic
networks display a power law shape: $P(k) \sim Ak^{-\alpha}$, where
typically $2<\alpha<3$. 
(iii) \textit{High average clustering coefficient}: in most real world networks, particularly social networks, nodes tend to create tightly knit groups with relatively high clustering coefficient.

\paragraph{Stochastic block model (SBM)} Let $n , k \in \mathbb{Z}^+$ denote the number of vertices and the communities, respectively; $p = (p_1, \dots , p_k)$ be a probability vector (the prior) on the $k$ communities, and $W \in  \{0,1\}^{k \times k}$ be a symmetric matrix of connectivity probabilities. The pair $(X,\Graph)$ is drawn
under the SBM$(n, p,W)$ if $X$ is an $n$-dimensional random vector with i.i.d. components
distributed under $p$, and $\mathcal{G(V,E)}$ is a simple graph where vertices $v$ and $u$ are
connected with probability $W_{X_v,X_u}$, independently of any other pairs. We define the community sets by $\Omega_i = \Omega_i (X) := \{v \in \mathcal{V} : X_v = i\}, i \in \{1,\dots ,k\}.$ 

Note that edges are independently but not identically distributed. Instead, they are conditionally independent,
i.e., conditioned on their groups, all edges are independent and for a given pair of groups $(i, j)$, they are i.i.d. Because each vertex in a
given group connects to all other vertices in the same way, vertices in the same community are said to be stochastically equivalent. The distribution of $(X, \Graph)$ for $x \in \{1, \dots, k\}^n$ is given by:
\begin{align*}
\prob\{ X=x \}:=& \prod\limits_{u=1}^n p_{x_u} = \prod\limits_{i=1}^k p_i^{|\Omega_i(x)|},\\
\prob\{ \mathcal{E}= y|X=x \}:= & \prod\limits_{1\leq u < v \leq n}  W_{x_u, x_v}^{y_{uv}}(1-W_{x_u,x_v})^{(1-y_{uv})}. 
\end{align*}

The law of large numbers implies that, almost surely, $\dfrac{1}{n}|\Omega_i| \rightarrow p_i $.

\textbf{Symmetric SBM (SSBM)} If the probability vector $p$ is uniform and $W$ has all diagonal entries equal to $q_{in}$ and all non-diagonal entries equal to $q_{out}$, then the SBM is said to be symmetric. We say $(X,\Graph)$ is drawn under the SSBM$(n, k, q_{in}, q_{out})$, where the community prior is $p = \{ 1/k \}^k$, and $X$ is drawn uniformly at random with the constraints $|\{ v \in \mathcal{V}: X_v=i \}|=n/k$. 
The case where $q_{in}> q_{out}$ is called assortative model. 
\section{Methods}
To design our four models we first generate a sequence of group sizes, and refer to the appendix for some of the detailed algorithms involved. Secondly, we produce the community structures according to the sequence of group sizes and add the interconnections among them in the four modalities of multi-group connectivity.

\subsection{Generating subgroup sizes}

In this section we describe an algorithm to generate relative subgroup sizes, and introduce the resulting properties of these subgroups. We compute a normalized sequence of group sizes with a heavy tail
distribution. We refer to Algorithm~\ref{algo:1} in the
appendix for a formal description based on pseudocode. Each subgroup is modeled as a connected dense Erd\H{o}s-R\'{e}nyi graph. For $\epsilon$
substantially smaller than 1 (we shall select it to be $10\%$), a subgroup
of size $i$ is the random graph $\Graph_{ER}(i,1-\eps)$. 

Each subgroup
of size $i$ and edge probability $1-\epsilon$ has the following properties:
\begin{enumerate}
\item connectivity threshold of $ t(i)=\dfrac{\ln(i)}{i}$, that is, for
  $1-\epsilon > t(i)$, $\Graph_{ER}$ is almost surely connected
  (almost any graph in the ensemble $\Graph_{ER}$ is connected);
\item $(1-\epsilon) \ddfrac{i(i-1)}{2}$ edges on average;
\item small average shortest path close to 1 and depending at most
  logarithmically on $i$;
\item binomial degree distribution: $P(k) = \binom{i-1}{k}(1-\epsilon)^k
  (\epsilon)^{i-1-k}$. Note that as $\epsilon$ decreases, the standard
  error becomes smaller and the distribution is more densely concentrated
  around the mean $(i-1)(1-\epsilon)$; and
\item large clustering coefficient close to 1 (conditioned on small $\epsilon$) and equal to $C=1-\epsilon$.
\end{enumerate}

\begin{figure}[h]
  \centering \includegraphics[width=.6\linewidth]{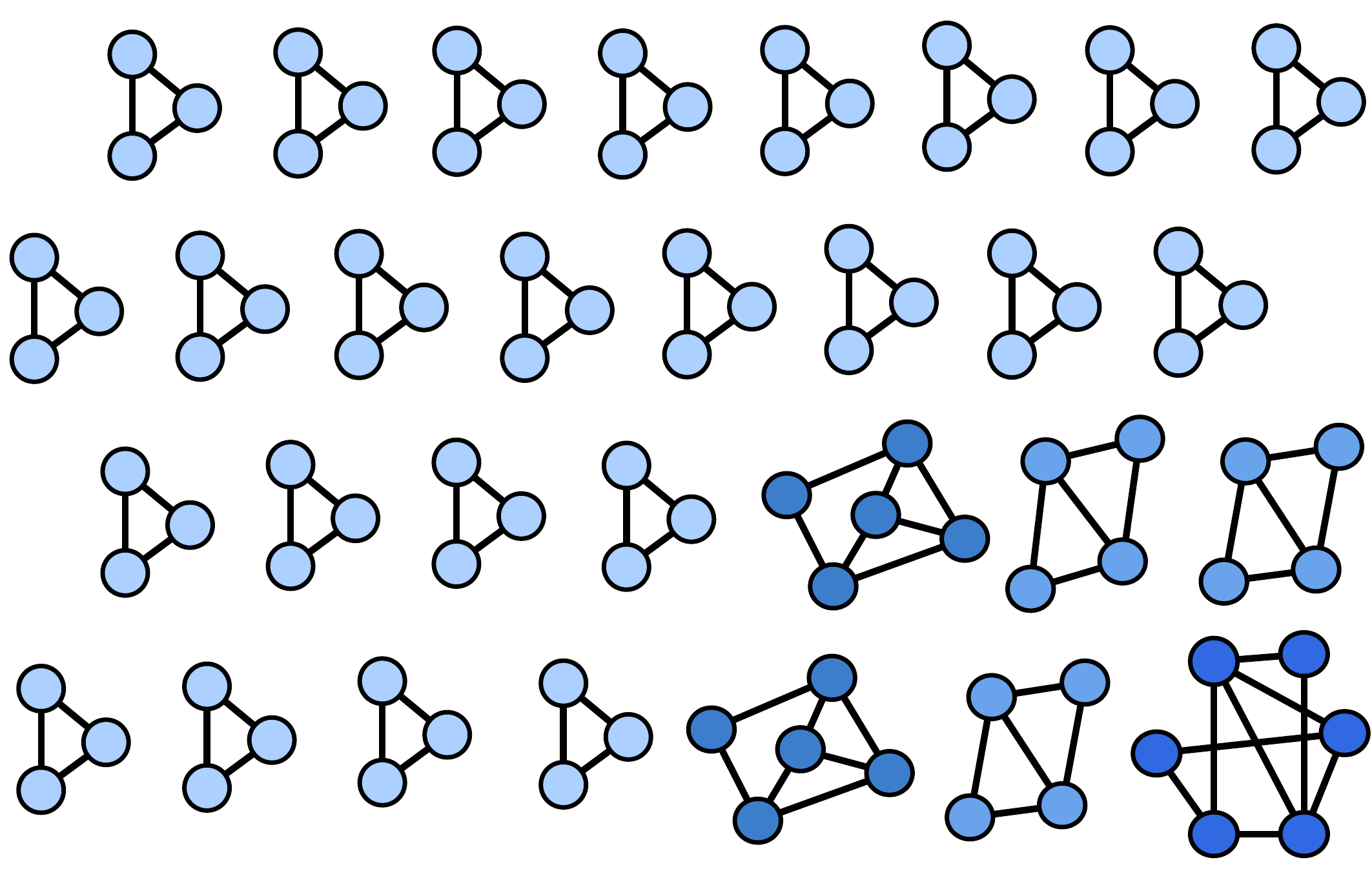}
  \caption{A collection of subgroups on 100 individuals.}
  \label{fig:collection-almost-cliques}
\end{figure}

Given a population of $n$ individuals, Algorithm~\ref{algo:1} generates a
sequence of relative subgroup sizes, such that, when interpreted as a disconnected graph, the collection of these subgroups exhibits a heavy tail degree distribution. An example of subgroup sizes generated by Algorithm~\ref{algo:1} is illustrated in
Fig.~\ref{fig:collection-almost-cliques}.

As part of Algorithm~\ref{algo:1}, we design the probability distribution
for the subgroup size $i$ to be proportional to~$\dfrac{1}{i^3}$. The
choice of exponent equal to $3$ is based on the following notes: first, in
order for $f(i)=\dfrac{k}{i^\alpha}$ and its mean to be well-defined, one
should have $\alpha \geq 2$; second, if one additionally requires the
distribution to have a finite variance, then $\alpha \geq 3$. With exponent
3, the outcome of each realization of the algorithm is a collection of
mostly small connected subgroups.

\subsection{Models of multi-group connectivity}
In this section we describe the algorithms that generate realizations of
the four multi-group connectivity modalities.

For three of the four modalities (bridges, edge bundles, and co-members), we connect the subgroups
  through a minimal set of pairwise coordination problems among
  them. Specifically, a minimal set of pairwise coordination problems is
  modeled through the notion of a random spanning tree among the
  subgroups. To define the generative algorithms for these three structures, we apply the notion of stochastic block models.

\subsubsection{Bridge connectivity model}

\begin{figure}[ht]
	\centering
	{\includegraphics[height=.4\linewidth]{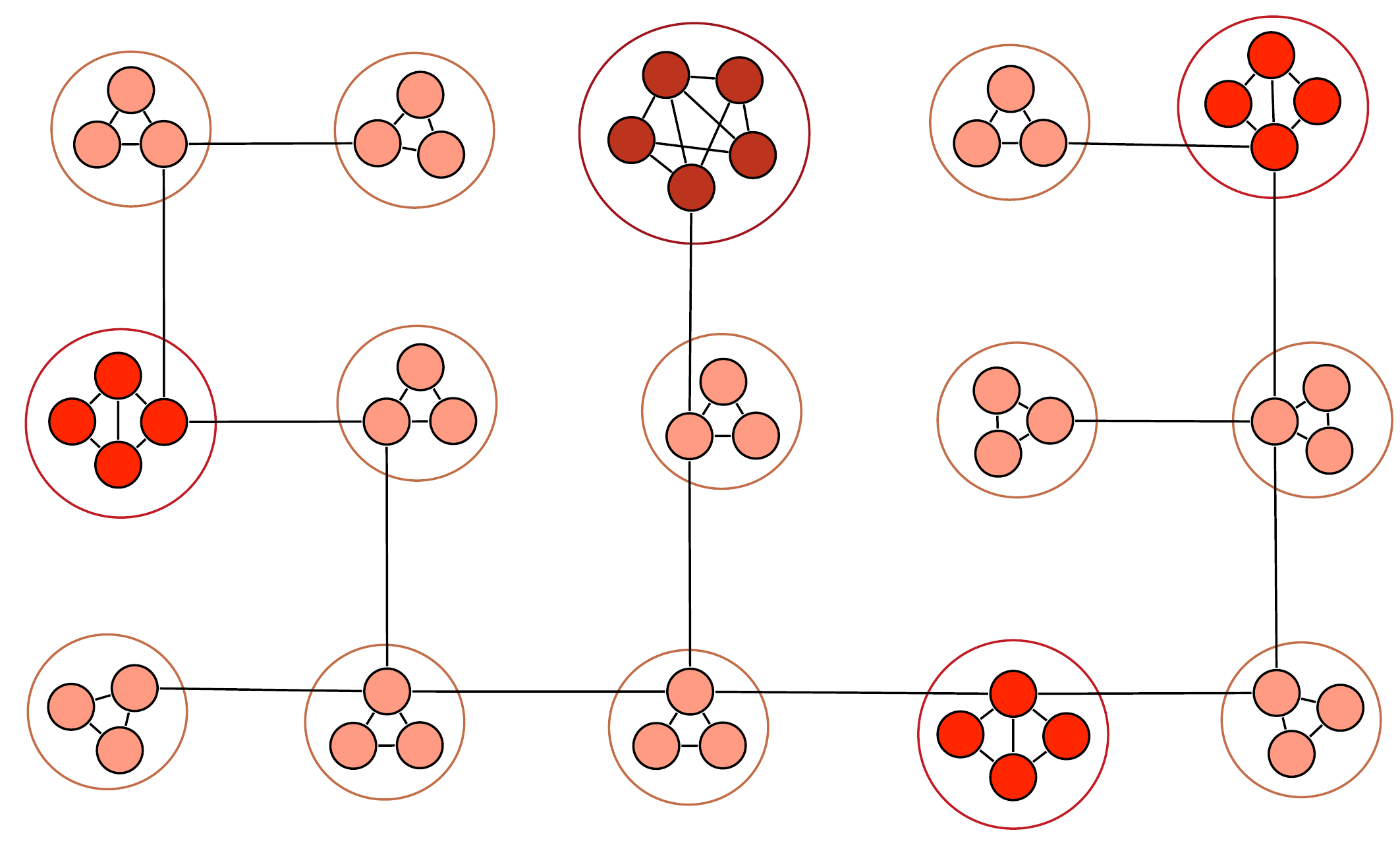}}
	\caption{Example of a network of 50 individuals in subgroups connected by bridges.}
	\label{fig:bridging-ties-2}
\end{figure}

Here we propose an algorithm to generate the bridge connected model. This structure can be modeled as a stochastic block model where the communities are connected through a uniform randomly generated spanning tree, and the interconnections are through precisely one node of each subgroup. We denote the edge set of this random tree with $\mathcal{E}_T$. The graph is drawn under the SBM$(n,p,W^B)$, conditioned under connectivity, where $p$ is calculated by Algorithm~\ref{algo:1}, and $W^B$ is given by: 

\begin{equation}
  W^{B}_{ij} =
    \begin{cases}
      1-\epsilon, & \quad \text{if $i=j$},\\
      \dfrac{1}{n^2 p_i p_j}=\dfrac{1}{s_is_j}, & \quad \text{if $i\neq j$ and $ij \in \mathcal{E}_T$},\\
      0, & \quad \text{otherwise},
    \end{cases}       
\end{equation}

where $s_i = |\Omega_i|$ denotes the size of group $i$, and $W^B$ contains a tree structure. 
Note that given an SBM, a node in community $i$ has $np_jW_{ij}$ neighbors in expectation in community $j$. We illustrate a realization of our algorithm in Fig.~\ref{fig:bridging-ties-2}.

\newpage
\subsubsection{Edge bundle connectivity model}
\begin{figure}[ht]
	\centering
	{\includegraphics[height=.4\linewidth]{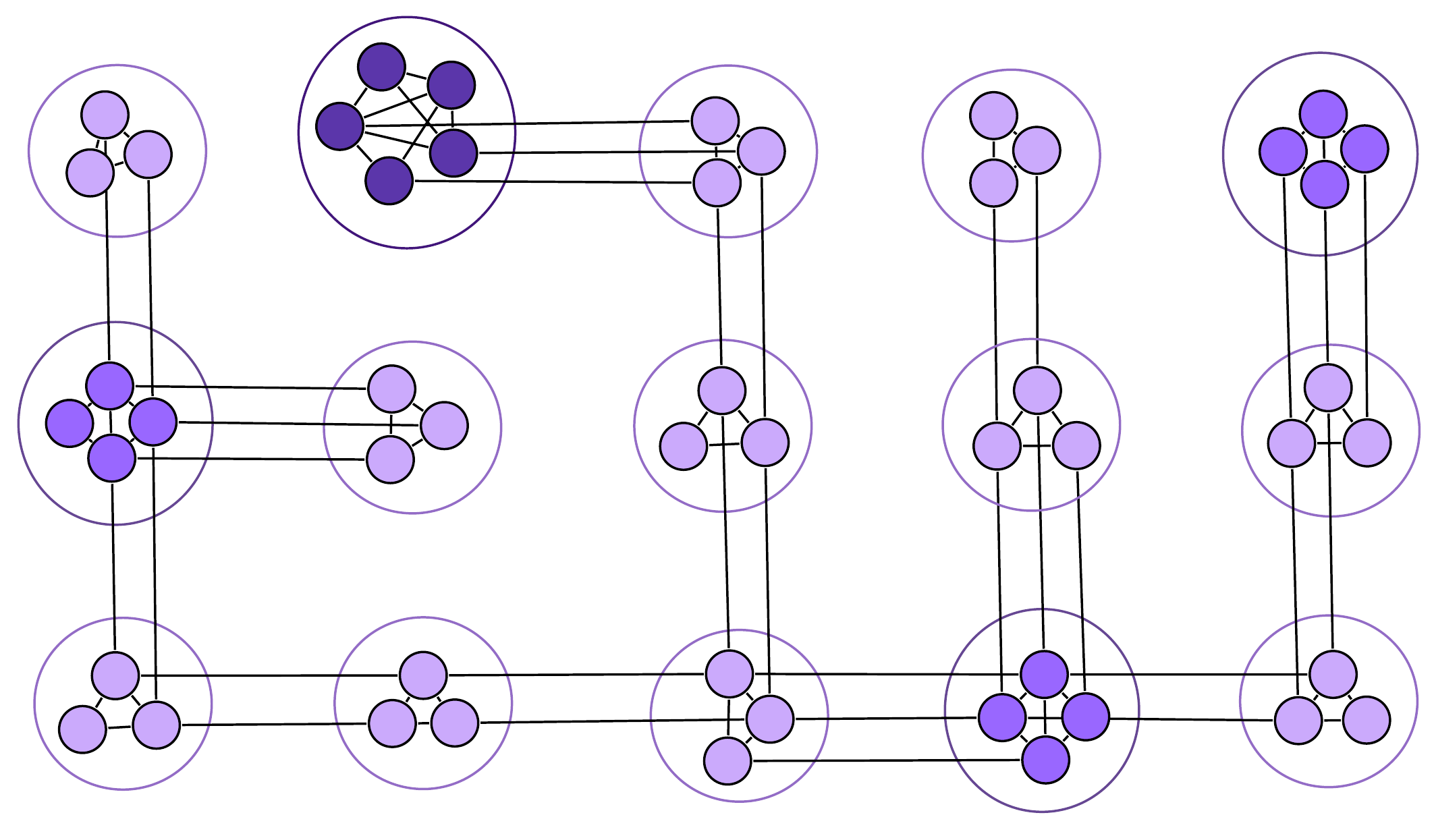}}
	\caption{Example of a network of 50 individuals in subgroups connected by bundles of edges.}
	\label{fig:red-ties-2}
\end{figure}

In this section we propose an algorithm to generate the edge bundle connectivity model. Again we apply a random spanning tree as the building block of the interconnections. Here, instead of adding a single edge as the basis of intergroup connectivity, we add multiple edges whose number grows with the size of the subgroups. We illustrate an algorithm realization in Fig.~\ref{fig:red-ties-2}.

We draw the graph under the SBM$(n,p,W^{EB})$, conditioned under redundant connectivity. Communities are connected through a uniform randomly generated spanning tree with edge set $\mathcal{E}_T$. The interconnections involve two or more nodes from neighboring subgroups. $p$ is calculated by Algorithm~\ref{algo:1}, and $W_{EB}$ is given by:

\begin{equation}
  W^{EB}_{ij} =
    \begin{cases}
      1-\epsilon, & \quad \text{if $i=j$},\\
	\dfrac{\alpha_{ij}}{n^2 p_i p_j}=\dfrac{\alpha_{ij}}{s_is_j}, & \quad \text{if $i\neq j$ and $ij \in \mathcal{E}_T$},\\
      0, & \quad \text{otherwise},
    \end{cases}       
\end{equation}
where $W^{EB}$ contains a tree structure, $\alpha_{ij} =\alpha_{ji} \geq 2$ for all $i,j$, and $\alpha_{ij}$ scales with $s_i s_j$.

\subsubsection{Co-membership connectivity model}
\begin{figure}[ht]
	\centering
	{\includegraphics[height=.375\linewidth]{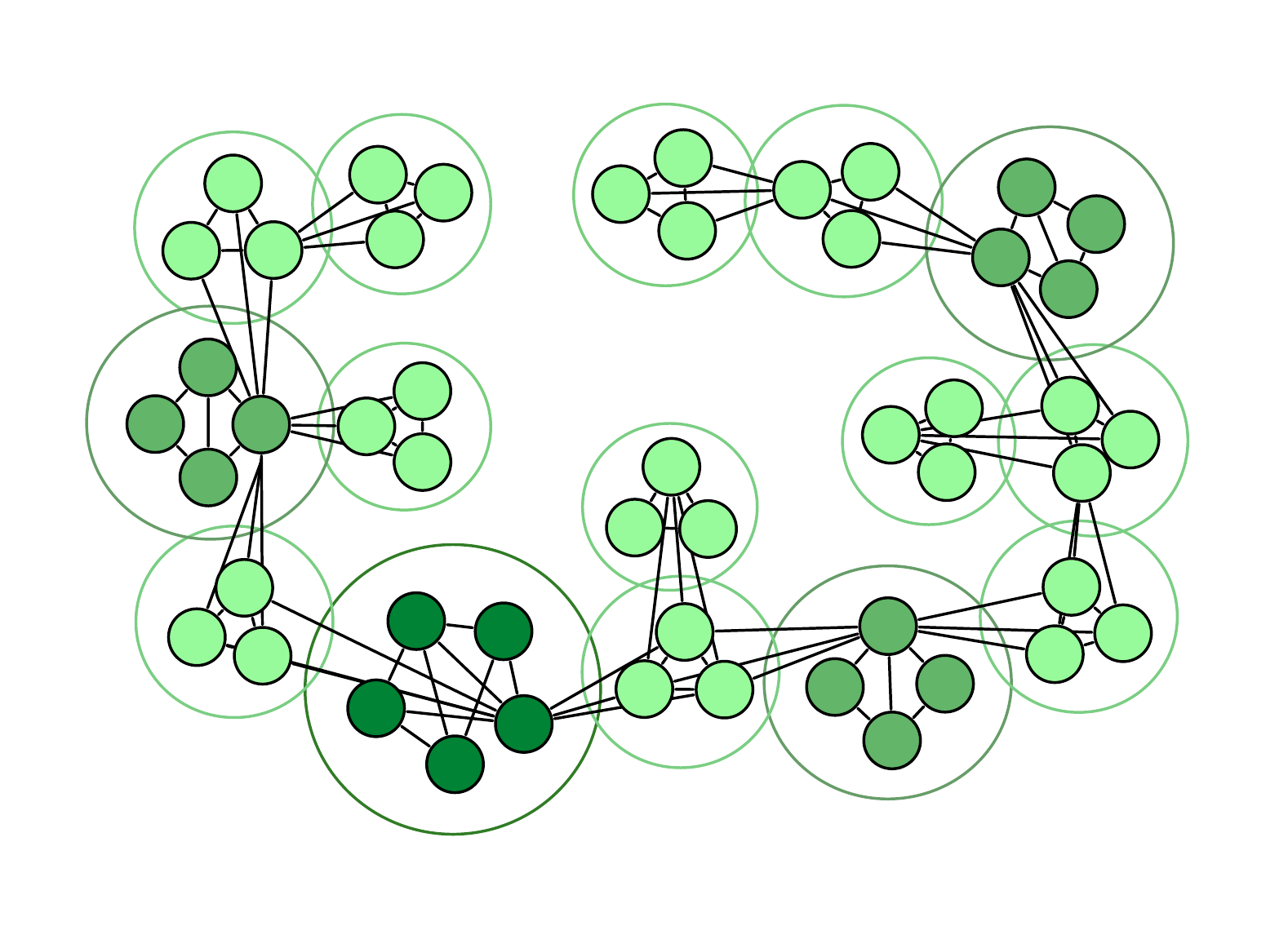}}\\
	\caption{Example of 50 individuals in a co-membership connected network.}
	\label{fig:comembership-2}
\end{figure}

In addition to the existence of a uniform random spanning tree over the subgroup, our co-membership connectivity model generation is conditioned under the following topological constraint: we
  consider each pair of connected subgroups, say $i$ and $j$, and select a
  fraction of edges in the complete bipartite graph over $i$ and $j$. For each of
  these selected edges, we randomly select one of the two individuals, say
  the individual in $i$, and we turn this individual into a member of the
  subgroup $j$ by adding edges from this individual to almost all members of $v$.We illustrate an algorithm realization in Fig.~\ref{fig:comembership-2}.
  
 The co-membership model can be generated as a realization of SBM$(n,p,W^{C})$, conditioned under the edge bundles initiated from a single node in one of the corresponding subgroups. Again $\mathcal{E}_T$ denotes the edge set of the random tree, $p$ is calculated by Algorithm~\ref{algo:1}, and $W^{C}$ is given by:

\begin{equation}
  W^{C}_{ij} =
    \begin{cases}
      1-\epsilon, & \quad \text{if $i=j$},\\
	\dfrac{\alpha_{ij}}{n^2 p_i p_j}=\dfrac{\alpha_{ij}}{s_is_j}, & \quad \text{if $i\neq j$ and $ij \in \mathcal{E}_T$},\\
      0, & \quad \text{otherwise},
    \end{cases}       
\end{equation}
where $W^{C}$ contains a tree structure, $\alpha_{ij} =\alpha_{ji} \geq 3$ for all $i,j$, and $\alpha_{ij}$ scales with either $s_i$ or $s_j$ ($\alpha_{ij} \approx s_i$ or $\alpha_{ij} \approx s_j$). 
  
\subsubsection{Liaison hierarchy connectivity model}
Here, applying Algorithm~\ref{algo:1} we first generate the subgroups as dense Erd\H{o}s-R\'{e}nyi graphs. Then we partition the subgroups into sets of 2 or 3, and (i) assign a liaison to each of sets and (ii) recursively assign a new liaison to groups of 2 or 3 liaisons until we reach the root at the top of the hierarchy. The resulting graph is a hierarchical tree structure with random branching factors of 2 and 3. A detailed description is provided in Algorithm~\ref{liaison} in the appendix,  and Fig.~\ref{fig:liaisons-2} illustrates a realization of this model.
\begin{figure}[ht]
	\centering
	{\includegraphics[height=.425\linewidth]{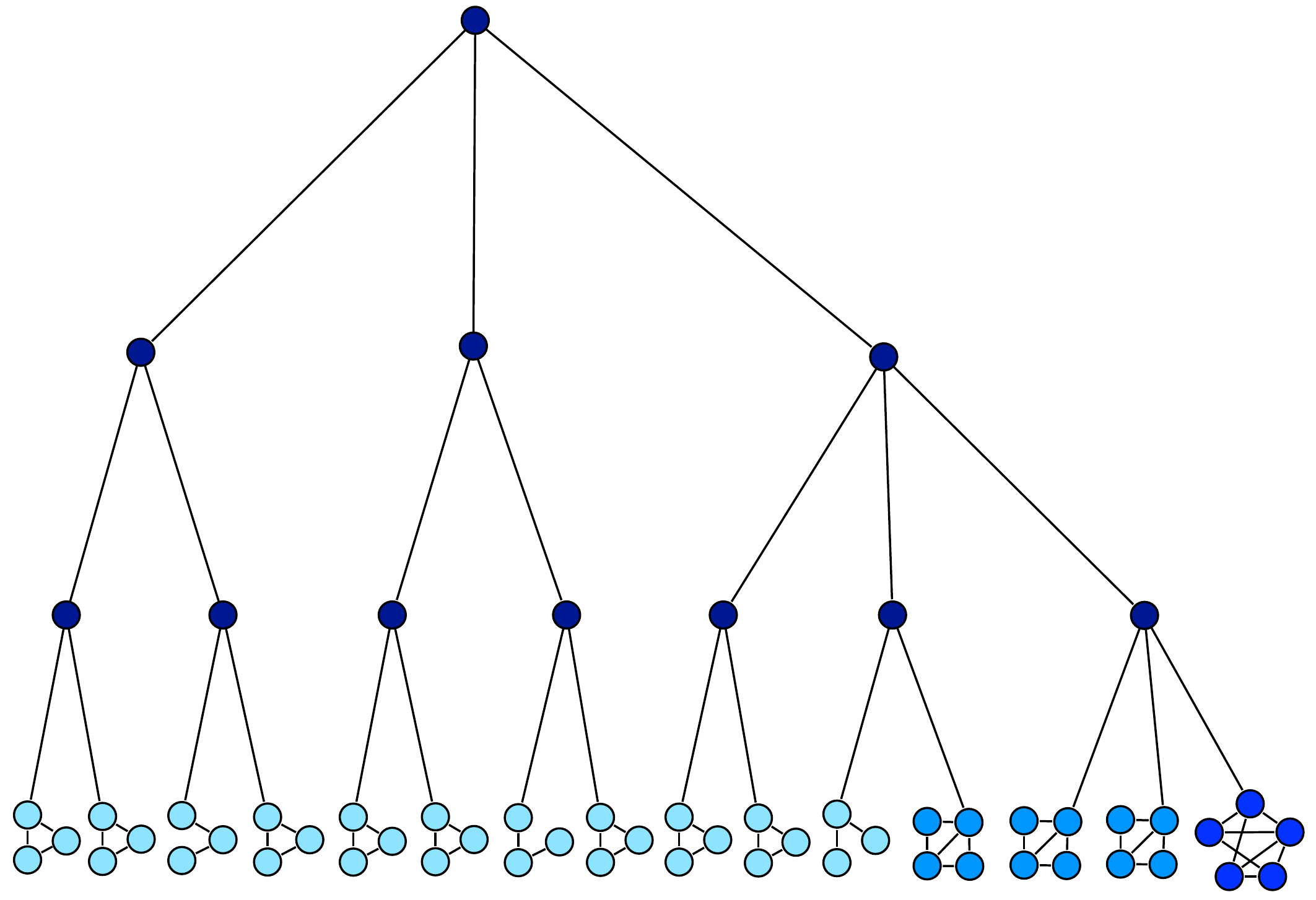}}
	\caption{Example of 50 individuals in subgroups joined by a liaison hierarchy generated by 
		Algorithm~\ref{liaison}.}
	\label{fig:liaisons-2}
\end{figure}

\section{Results} \label{comparison-structurel}
Realistic networks are usually not exclusively based on a single modality of subgroup connectivity. Our comparative analysis of connectivity modalities is oriented to the question of the implications of a shift away from one modality toward another modality. For example, a modality shift from a liaison hierarchy toward direct bridges among subgroups, or from bridges among subgroups to intergroup edge bundles, or from intergroup edge bundles to co-memberships.        

\begin{figure}[ht]
\begin{center} 
\subfloat[Plot of average shortest path]{\includegraphics[scale = 0.4]{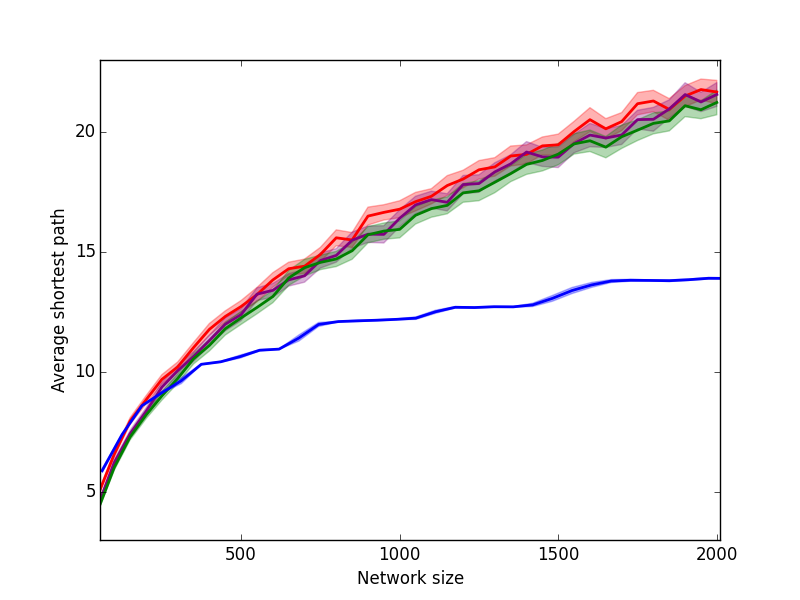}\label{fig:ave-shortest-dist}}\\
\subfloat[Plot of average degree]{\includegraphics[scale = 0.4]{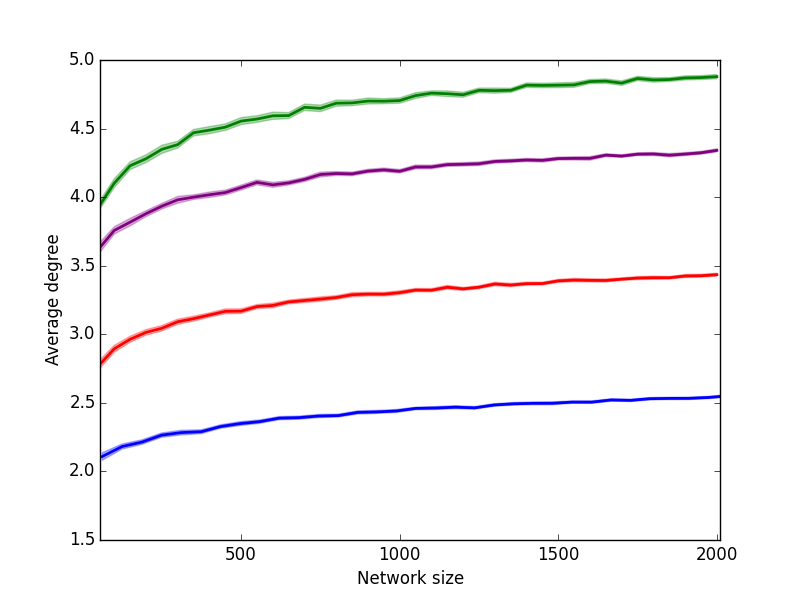}\label{fig:ave-degree}}
\caption{\small In each plot red refers to the bridge model, purple to the edge bundle model, green to the co-membership model, and blue to the liaison hierarchy model}\label{fig:structure-comp}
\end{center}
\end{figure}

In Fig.~\ref{fig:structure-comp} we present a comparison of the average
shortest paths and average degrees of our generated networks as a function
of network size for each of the four multi-group connectivity
modalities. Each sample point on the curves is based on 100 realizations on
networks with sizes that increase in step sizes of 50 up to 2,000 nodes. In
analyses that increase the sample point size to 1,500 over a range of sizes
up to 500, there is no marked change in the trajectories. In general, the
confidence interval bands are narrow. Here, and elsewhere, red refers to
the bridge model, purple to the edge bundle model, green to the
co-membership model, and blue to the liaison hierarchy model. Fig.~\ref{fig:ave-shortest-dist} shows that the liaison hierarchy
increasingly distinguishes itself from the three modalities as network size
increases. Its displayed trajectory is conditional on the liaison structure
design. Average shortest paths are insensitive to redundancies. Hence, the
lack of distinctions among the other three modalities is not
surprising. Fig.~\ref{fig:ave-degree} shows that the four modalities are
systematically ordered with respect to their average degrees:
(co-membership) $>$ (edge-bundle) $>$ (bridge) $>$ (liaison) with respect
to their average degrees.

\subsubsection{Spectral radius and propagation processes} \label{propagation-processes}

Propagation phenomena appear in various disciplines, such as spread of
infectious diseases, transmission of information, diffusion of innovations,
cascading failures in power grids, and spread of wild-fires in
forests. Based on the application, the objective can vary from avoiding
epidemic outbreaks and eradicating the disease in a population to
facilitating the spread of an ideology or product over a network in
marketing campaigns. In this subsection we provide a comparison of the
system behavior under the simple and well-studied epidemic models proposed
in the literature for our four proposed network models.

Let $x(t)=\big(x_1(t),\dots,x_n(t)\big)^{\top}$ denote the
  infection probabilities of each node at time $t$ and $A\in \real^{n\times
    n}$ denote the adjacency matrix of the contact graph. Let $\beta>0$ be
  the infection rate, and $\gamma>0$ be the \emph{recovery rate} to the
  susceptible state.  Then the linearization of the \emph{SI
    (Susceptible-Infected) and SIS (Susceptible-Infected-Susceptible)
    network propagation models} about the no-infection equilibrium point
  $\vectorzeros[n]$ on a weighted digraph are given by, respectively,
\begin{align}
  \dot{x} &= \beta A x, \label{def:network-SI-vector}\\
 \dot{x}  &= (\beta A - \gamma I_n)x.  \label{def:network-SIS}
\end{align}
The following results are well known (see the classic
works~\cite{AL-JAY:76,LJSA:94,YW-DC-CW-CF:03} and the recent
review~\cite{WM-SM-SZ-FB:16f}).  In the SI model the epidemic initially
experiences exponential growth with rate $\beta \lmax$. In the SIS model
near the onset of an epidemic outbreak, the exponential growth rate is
$\beta\lmax-\gamma$ and the outbreak tends to align with the dominant
eigenvector.

\begin{figure}[ht]
	\centering
	\includegraphics[scale=0.425]{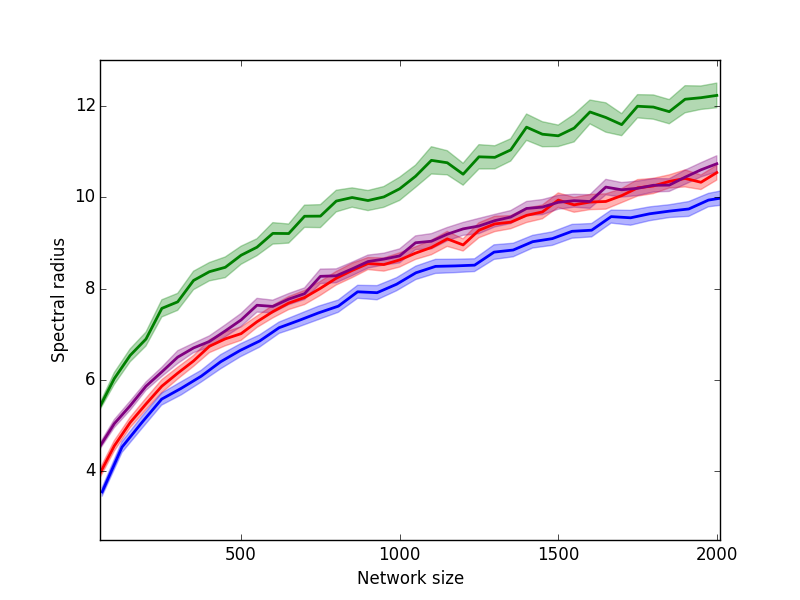}
	\caption{Plot of spectral radius}
	\label{fig:spectral-rad}
\end{figure}

In Fig.~\ref{fig:spectral-rad} we plot the spectral radius of the networks as a function of network size 50-2,000 for the four  models. In Table \ref{nonlinear:sr} we evaluate the differences among these curves controlling a network's size (N), average degree (Degree), and (0,1) indicator variables for the edge-bundle, co-membership, and liaison modalities with the bridge modality taken as the baseline. Similar findings were obtained with 660K observations on a reduced range of network sizes 24-655. The average degree of a network has a positive effect on the speed of viral propagation. Controlling for network size and average degree, relative to the propagation speeds in the bridge modality, propagation speeds in the edge-bundle and liaison modalities are greater and those of the co-membership modality are less. The elevated curve for co-membership modality in Fig.~\ref{fig:spectral-rad} is based on its systematically higher average degrees.         

\begin{table}[ht]
	\centering
	\caption{Nonlinear regression results for spectral radius, controlling for network size and average degree, and indicator variables for the connectivity modalities with bridge modality as baseline (15,200 networks, R$^2$= 0.833)}
        \resizebox{\linewidth}{!}{
	\begin{tabular}{ p{3cm} p{3cm} p{3cm} c } 
		\hline \hline
	       & coeff.  & s.e. & p-value \\ 
		\hline
		Constant &       -2.8103 &      0.12916    &   $<$.0001  \\
		N        &       0.0038847&    5.2749e-05  &   $<$.0001 \\
		Degree   &       5.0734 &     0.088421     &   $<$.0001 \\
		Edge-bundle &    0.2012 &     0.019035     &   $<$.0001 \\
		Co-membership &  -1.8589&      0.062881    &   $<$.0001 \\
		Liaison &        0.24287&      0.023229    &   $<$.0001 \\
		N$^2$ &          -8.3082e-07&    2.1709e-08&   $<$.0001 \\
	   \hline \hline
	\end{tabular}}
    \label{nonlinear:sr} 
\end{table}

\subsubsection{Time to convergence in influence processes generating consensus with distributed linear averaging} \label{noiseless-consensusl}
Consensus algorithms play an important role in many multi-agent
systems. They are usually defined as in French-DeGroot discrete-time
averaging recursion

\begin{equation}\label{eqn:consensus}
 x(t+1)= W x(t).
\end{equation}
\noindent where $W$ is row stochastic and $x(t) \in \mathbb{R}^n$ is the vector of individuals' opinions at time $t$. For primitive stochastic matrices the solution to \eqref{eqn:consensus} satisfies 
\begin{equation}
\lim_{k \rightarrow \infty} x(k) = \big(v^T x(0)\big)\vectorones[n] 
\end{equation}
where $v$ is the left dominant eigenvector of $W$ satisfying $v_1+\dots
+v_n =1$. Convergence time to consensus may be defined as
$\subscr{\tau}{asym} = \ddfrac{1}{\log(1/\subscr{r}{asym})}$ and it gives
the asymptotic number of steps for the error to decrease by the factor
$1/e$, where $\subscr{r}{asym}$ denotes the asymptotic convergence factor.
It is well known, e.g., see~\cite{FB:18}, Chapter 10, that convergence to
consensus is exponentially fast as $\rho_2^t$, where $\rho_2$ is the second
largest eigenvalue of $W$ in magnitude.  We construct $W$ from $A$ as
follows:
\begin{equation}\label{eqn:equal-neighboring}
  W=(D+I_n)^{-1} (A+I_n),
\end{equation}
\noindent where $D=\diag(A \vectorones[n])$ denotes the diagonal matrix of all the
nodes' out-degrees, with $d_{ii} = \sum_{j=1}^{n}a_{ij}$ $\forall i$. Equation~\ref{eqn:equal-neighboring} gives positive weights $w_{ii}$ that are equal to the $w_{ij}$ weights of $i$'s neighbors in $A$. 

\begin{figure}[ht]
  \centering
  \includegraphics[scale=0.425]{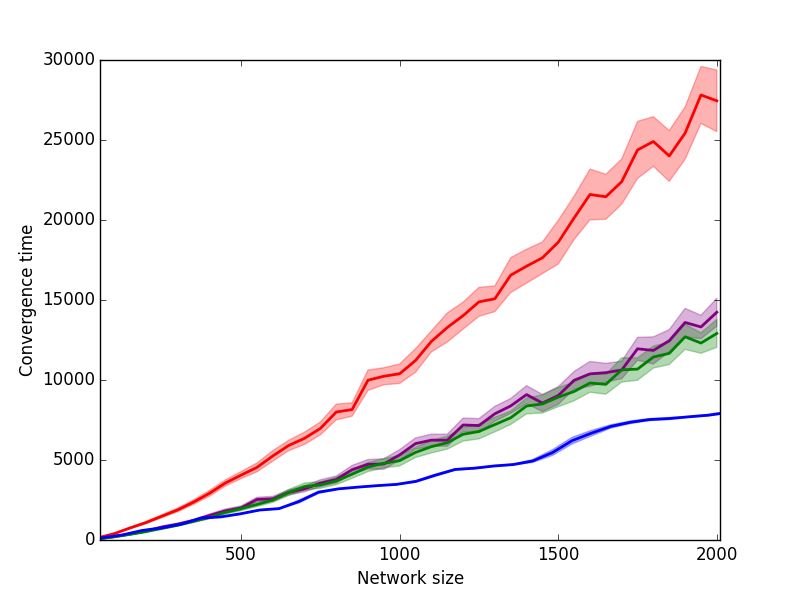}
  \caption{Plot of convergence time for the four network models with equal neighboring weights}
  \label{fig:convergence-time}
\end{figure}

In Fig.~\ref{fig:convergence-time}, we plot the average convergence-times of the networks as a function of network size 50-2,000 for the four  models.  In Table \ref{nonlinear:ct} we evaluate the differences among these curves controlling a network's size (N), average degree (Degree), and (0,1) indicator variables for the edge-bundle, co-membership, and liaison modalities with the bridge modality taken as the baseline. Similar findings were obtained with 660K observations on a reduced range of network sizes 24-655. The convergence times of the bridge modality are larger than those of the three other modalities, and the liaison modality has the fastest convergence times. Higher average degrees lower times to convergence. Controlling for network size and average degree, the convergence times of the edge-bundle modality are faster than those of the co-membership modality. 

\begin{table}[ht]
	\centering
	\caption{Nonlinear regression results for convergence time, controlling for network size and average degree, and indicator variables for the connectivity modalities with bridge modality as baseline (15,200 networks, R$^2$= 0.637)}
    \resizebox{\linewidth}{!}{
	\begin{tabular}{p{3cm} p{3cm} p{3cm} c} 
		\hline \hline
		& coeff.  & s.e. & p-value \\ 
		\hline
		Constant &         6991.1   &     590.12  &       $<$.0001     \\
		N        &         11.558 &      0.241  &       $<$.0001     \\
		Degree    &       -3300.9 &      403.98 &    $<$.0001     \\
		Edge-bundle&      -4652.9 &     86.969  &$<$.0001     \\
		Co-membership&     -2993.9 &     287.29 & $<$.0001     \\
		Liaison       &    -7790.3 &    106.13  &  $<$.0001     \\
		N$^2$        &    -0.0018372 &  9.9184e-05 &    $<$.0001     \\
	\hline \hline
	\end{tabular}}
	\label{nonlinear:ct} 
\end{table}

\subsubsection{Consensus processes subject to white Gaussian noise}
The general form of a French-DeGroot influence process with white Gaussian
noise is:
\begin{equation} \label{noisy-concensus}
x(t+1)= W x(t) + e(t)
\end{equation}
\noindent where $e(t)$ is a random vector with zero mean and covariance $\Sigma_e$ having independent entries. In the presence of noise, the states of the agents will be brought close to each other, but will not fully align to exact consensus. The resulting noisy consensus is referred to as \textit{persistent disagreement}. For strongly connected and aperiodic graphs, the consensus dynamics \eqref{eqn:consensus} correspond to an irreducible and aperiodic Markov chain. The matrix $W$ then corresponds to the transition probability matrix and its normalized left dominant eigenvector $\pi$
corresponds to the stationary distribution vector of the chain. The results on the steady-state disagreement by Jadbabaie \& Olshevsky~\citep{AJ-AO:16} apply to reversible Markov chains which with the choice of weights on our adjacency matrix will be met.  For the Markov chain with reversible transition matrix $W$ and with uncorrelated noise, the mean square asymptotic error $\delta_{ss}$ can be measured by:
\begin{equation}\label{thm:delta_ss}
\delta_{ss} = \pi^T H D_{\pi} \Sigma_e D_{\pi} \vectorones[n], 
\end{equation}
\noindent where $D_{\pi}=\diag(\pi)$, and $H$ is the matrix of hitting
times for the Markov chain. The algorithm by Kemeny \& Snell~\citep{JGK-JLS:76} is applied
to compute $H$.

\begin{figure}[ht]
  \centering
  \includegraphics[scale=0.425]{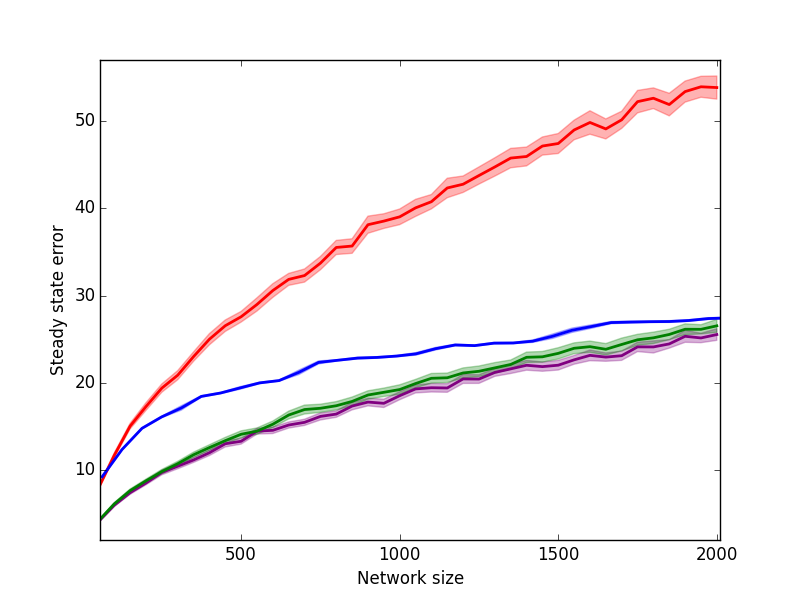}
  \caption{Plot of $\delta_{ss}$ of graph $A$ for the four network models with equal neighboring weights}
  \label{fig:ss-mean-square-dev}
\end{figure}

In Fig.~\ref{fig:ss-mean-square-dev}, we plot the steady-state mean deviation from consensus, given by~\eqref{thm:delta_ss}, on the networks as a function of network size 50-2,000 for the four  models.  In Table \ref{nonlinear:dev} we evaluate the differences among these curves controlling a network's size (N), average degree (Degree), and (0,1) indicator variables for the edge-bundle, co-membership, and liaison modalities with the bridge modality taken as the baseline. Similar findings were obtained with 660K observations on a reduced range of network sizes 24-655. 
The steady-state mean deviations for the bridge modality are larger than those of the three other modalities. Higher average degrees lower steady-state mean deviations from consensus. Although the modalities have distinguishable effects, again we note that average degree differences are ``boiled into'' the modality models, so that when average degree is controlled, the relative ordering of modalities is altered. The edge-bundle and liaison modalities have greater noise reduction properties than the co-membership modality.

\begin{table}[ht]
	\centering
	\caption{Nonlinear regression results for steady-state mean deviation from consensus, controlling for network size and average degree, and indicator variables for the connectivity modalities with bridge modality as baseline (15200 networks, R$^2$=  0.768)}
     \resizebox{\linewidth}{!}{
	\begin{tabular}{ p{3cm} p{3cm} p{3cm} c } 
		\hline \hline
		& coeff.  & s.e. & p-value \\ 
		\hline
		Constant   &      30.894   &    0.84085        & $<$.0001 \\
		N            &    0.031003   &  0.0003434        & $<$.0001 \\
		Degree            &     -9.1312   &    0.57562        & $<$.0001 \\
		Edge-bundle          &     -16.807   &    0.12392        & $<$.0001 \\
		Co-membership            &     -10.502   &   0.40935       & $<$.0001 \\
		Liaison           &     -14.052   &    0.15122       & $<$.0001 \\
		N$^2$             & -9.1291e-06   & 1.4133e-07       & $<$.0001 \\
	\hline \hline
	\end{tabular}}
	\label{nonlinear:dev} 
\end{table}

\section{Discussion}
In this article, we have proposed simple, synergistic, and stochastic
algorithms to generate four modalities of multi-group connectivity and have
compared their implications. These algorithms are a variant of what is known as planted partition or stochastic block models, under some further topological constraints including that the intergroup connectivity is shaped by an underlying tree. Models 1-3 are nested in the following sense: for appropriate parameters, 1) graphs generated by the bridge connectivity structure are subgraphs of those generated by the edge bundles, and 2) graphs generated by the edge bundle connectivity structure could be subgraphs of those generated by the co-membership. However, moving from the edge bundles to co-memberships, we introduce an additional constraint, that is, edge bundles of the spanning tree are initiated from the same node in one of neighboring subgroups in the co-membership model. The work touches on two central traditions in network analysis: models of network structure and models of dynamical processes that unfold on networks composed of multiple small groups with
dense within-group edges. In a connected network, any two such groups might
be intersecting (with one or more individuals who are members of both) or
disjoint. Two disjoint subgroups may be linked by a bridge, or by multiple
edges, or by individuals who are not members of any dense group. We
consider networks that can be strictly characterized in terms of one of
these types of inter-group connectivity. The touchstone for our analysis is
the work that has been conducted on multiple-group connectivity based on
bridges. Here we elaborate the analysis with a comparison of implications
of group-connectivity based on (i) a minimal set of bridges, (ii) a minimal
block-model structure in which pairs of groups are linked by multiple
edges, (iii) a minimal set of group membership intersections, and (iv) a
hierarchical tree of group-independent agents (intermediary liaisons.) No
doubt there are many ways to construct realizations of each type of
connectivity. No doubt there are many process metrics that might be
examined. We compare structures in terms of network process metrics. We
focus on metrics of two processes---epidemic propagation and consensus
formation.  These metrics are sensitive to network topology. We emphasize that the results of these comparative analyses are not merely due to the different numbers of links being added
to isolated clusters. The regression results controlling for the network sizes and average node degrees affirm this claim. We constrain
topology to four broad classes of markedly non-random clustered
networks. Our contribution is to show the feasibility of a principled
approach to a comparative analysis that we believe is currently lacking
with respect to these distinguishable topological classes.

Our findings on the speed of viral propagation show that the speeds differ depending on the form of multi-group connectivity.   The average degree of a network has a positive effect on the speed of viral propagation. If the average degrees differences, shown in Fig.~\ref{fig:ave-degree}, are characteristic features of the modalities, then Fig.~\ref{fig:spectral-rad} shows the net effect of each modality. Controlling for network size and average degree, our regression analysis in Table~\ref{nonlinear:sr} evaluates the independent contributions of average degree and modality type. If it were possible to construct modality types with identical average degrees, then the regression results suggest that the bridge, edge-bundle, and liaison modalities do not substantially differ in their speeds of viral propagation, and that the co-membership modality dampens the speed of viral propagation.  

Our findings on the times to convergence to consensus show that convergence times differ depending on the form of multi-group connectivity. The average degree of a network has a negative effect on convergence times, that is, higher average degrees are associated with faster convergence to consensus. If the average degrees differences, shown in Fig.~\ref{fig:ave-degree}, are characteristic features of the modalities, then Fig.~\ref{fig:convergence-time} shows the net effect of each modality. The bridge modality has slower convergence times than all other modalities. If it were possible to construct modality types with identical average degrees, then the regression results in Table~\ref{nonlinear:ct} suggest somewhat similar results. As in Fig.~\ref{fig:convergence-time}, the convergence times in the bridge modality are greater than all other modalities, and the liaison modality has the fastest convergence times. The regression on the edge-bundle and co-membership modalities indicates that, for a given average degree and network size, convergence is faster for edge-bundles than co-membership modalities. 
   
Finally, our findings for levels of steady-state stochastic deviations from consensus in the presence of noise show that the mean deviations differ depending on the form of multi-group connectivity. The average degree of a network has a negative effect on steady-deviation, that is, higher average degrees are associated with smaller deviations (more reduction of noise).  If the average degrees differences, shown in Fig.~\ref{fig:ave-degree}, are characteristic features of the modalities, then Fig.~ \ref{fig:ss-mean-square-dev} shows the net effect of each modality. The bridge modality has greater deviations (less reduction of noise) than all other modalities. If it were possible to construct modality types with identical average degrees, then the regression results in Table~\ref{nonlinear:dev} suggest somewhat similar results. As in Fig.~\ref{fig:convergence-time}, the levels of noise reduction in the bridge modality are less than in all other modalities. The regression on the edge-bundle, co-membership and liaison modalities indicate that edge-bundles are associated with the greatest reduction of noise. 

The important caveat on our findings is that they are conditional on
positions taken in the models with which we generated realizations of each
modality; see Algorithms~\ref{algo:1}-\ref{algo:greedy+incremental} in the
appendix. In addition, although it is reasonable that differences of
average degree are associated with different modalities, we have not
derived bounds on average degree for each modality (this may be an
intractable problem). Furthermore, our analysis of multi-group connectivity
modalities involves a uniform modality, whereas real networks with multiple
subgroups are likely to be connected with mixed modalities including
instances of bridges, edge-bundles, co-memberships, and liaison nodes who
are not members of any group. We believe that these obvious limitations are
out-weighted by the insights obtained from an analysis of artificial
network topologies with controllable features. In the set of
  findings of this paper, we were particularly struck by (1) the
  implications on network process metrics of the social cohesion entailed
  in edge-bundle and co-membership modalities of multi-group connectivity,
  and (2) by the strong effects on process metrics of network differences
  of average degree arising from the multiple modalities.
  
  An interesting future research direction is to propose sufficiently predictive indicators that enable one to categorize an arbitrary graph into any of the four connectivity structures discussed in this paper. In other words, we are interested in the following question: ``given an empirically observed graph, can one provide a computationally efficient algorithm to identify subgroups and classify them into these different connectivity structures?'' We find the results on the following literature relevant: recovery of the communities in the prolific community detection literature~\cite{MEJN-MG:04, SF:10}, graph clustering~\cite{SES:07}, and graph modularity~\cite{MEJN:06}.  Stochastic block models are widely recognized generative models for community detection and clustering in graphs and they provide a ground truth for identifying subgroups. \cite{EA:17} surveys recent developments for necessary and sufficient conditions for community recovery and community detection in SBMs.
  
 \section{Acknowledgments}
The authors thank professor Ambuj K. Singh for his valuable comments and suggestions. This material is based upon work supported by, or in part by, the U.S.~Army Research Laboratory and the U.S.~Army Research Office under grant numbers W911NF-15-1-0577.

\clearpage
\newpage

\appendix 
\section*{Appendix: Algorithm specifications}
In this appendix we present a detailed pseudocode description for three
relevant algorithms. Specifically, we present pseudocode for generating relative
subgroup sizes, for the sub-problem of generating a sequence of realizations of a random variable subject to a fixed sum, and for the liaison generative model. \label{appendix}

{\SetAlgoNoLine
\begin{algorithm} 
	\caption{\textbf{Generating sequence of relative subgroup sizes}\label{algo:1}}
	\KwIn{$n=$ number of nodes}
	\Parameter{
		$\alpha = 3$ exponent of power law \\
	}
	\KwOut{sequence of relative subgroup sizes $p$}  
	\Indentp{.5em}
	define a random variable $X$ taking values over $\{3, 4, \dots, n
	\}$, with probability mass function $P[X=x] \propto {1/x^3}$ to denote the random size of the subgroups
	
	invoke Algorithm~\ref{algo:greedy+incremental} to incrementally and greedily generate a sequence
	of realizations for $X$, denoted by $\{S_1,\dots,S_k\}$, satisfying the
	constraint $S_1+\dots+S_k=n$
	
	\For {$i=1:k$}{ 
	\Indentp{1em} $p_i \leftarrow S_i/n$}
	
	\Return{p}  
\end{algorithm}}

{\SetAlgoNoLine
\begin{algorithm}
\caption{\textbf{Generating a sequence of realizations of a given random variable with fixed sum}}\label{algo:greedy+incremental}
\KwIn{a discrete variable $X$ taking values in $\{x_{\min},\dots,x_{\max}\}$ with given pmf, number: $n$}

\KwOut{$S = \{ S_1,\dots,S_k\}$ a sequence of realizations of $X$,
  adjusted in a greedy incremental way such that $S_0+\dots+S_k=n$}
\Indentp{.5em}
$S \leftarrow \{\}, \subscr{n}{tmp}\leftarrow n$

\While{$\subscr{n}{tmp} \geq x_{\min}$}{
\Indentp{1em}
$\bar{x}\leftarrow$ realization of $X$

  \If{$\bar{x} \leq  \subscr{n}{tmp}$}{
  \Indentp{1em}
  $S\leftarrow S \cup \{ \bar{x} \}$, 
  
  $\subscr{n}{tmp} \leftarrow \subscr{n}{tmp}- \bar{x}$
}}

\For{$i=1:\subscr{n}{tmp}$}{
\Indentp{1em}
randomly select an number $S^*$ in the
  sequence~$S$ satisfying $S^*< x_{\max}$
  
  $S^*\leftarrow S^*+1$
}

\Return{S} 
\end{algorithm}}

{\SetAlgoNoLine
\begin{algorithm}
	\caption{\textbf{Liaison hierarchy connectivity}}\label{liaison}
	\KwIn{collection of subgroups generated using Algorithm~\ref{algo:1}}
	\Parameter{branching factor of each liaison = 2 or 3 }
	\KwOut{graph composed of subgroups plus hierarchy interconnections}
	\Indentp{.5em}
	define a random variable $L$ taking values over $\{2, 3\}$, with pmf
	$P[L=l] \propto {1/l^3}$ to denote the random branching factor of
	liaisons

	$n_l \leftarrow$ no. of subgroups
	
	\While{$n_l>1$}{
		\Indentp{1em}
		invoke Algorithm~\ref{algo:greedy+incremental} to
		generate a sequence of realizations for $L$, denoted by
		$\{S_1,\dots,S_k\}$, satisfying the constraint $S_1+\dots+S_k=n_l$
		
		\For {$i=1:k$}{
	    \Indentp{1em}		
		generate a liaison with branching factor $S_i$
			
			incrementally connect the liaison to $S_i$ unattended subgroups, if any exist, or unattended liaisons, after attending to all subgroups }
		
		$n_l \leftarrow k$
	}
	assign one liaison to the top of the hierarchy
	
	\Return{hierarchical tree with the subgroups as the leaves
	}
	
\end{algorithm}}

\clearpage

\section*{References}
\bibliographystyle{plainnat}
\bibliography{alias,Main,FB}

\end{document}